\documentstyle[pjw_cite3,psfig,graphicx]{mn}

\def\la{\mathrel{\hbox{\rlap{\hbox{\lower4pt\hbox{$\sim$}}}\hbox{$<$}}}}
\def\ga{\mathrel{\hbox{\rlap{\hbox{\lower4pt\hbox{$\sim$}}}\hbox{$>$}}}}

\title[X-ray observations of 4 Draconis]{X-ray observations of 4 Draconis: symbiotic binary or cataclysmic triple? }

\author[P.J. Wheatley, K. Mukai \& D. de Martino]
	{Peter J. Wheatley$^1$, Koji Mukai$^{2,3}$ \& Domitilla de Martino$^{4}$\\
	 $^1$Department of Physics and Astronomy, University of Leicester, 
	     University Road, Leicester LE1 7RH \\
	 $^2$Laboratory for High Energy Astrophysics, NASA/GSFC, Code 662, 
             Greenbelt, MD\,20771, USA \\
	 $^3$Universities Space Research Association\\
	 $^4$INAF-Osservatorio Astronomico de Capodimonte, via Moiariello n.\,16, 
          I-80131 Napoli, Italy}

\begin{document}

\maketitle

\begin{abstract}
We present the first X-ray observations of the 4~Draconis system,
consisting of an M3\,III giant with a hot ultraviolet companion. 
It has been claimed that the companion is itself an AM~Her-type binary 
system, an identification that places strong constraints on the
evolution of cataclysmic variables.  We find that the X-ray properties
of 4~Draconis are consistent with the presence of an accreting white
dwarf, but not consistent with the presence of an AM~Her system. 
We conclude that 4~Dra is therefore most-likely a symbiotic binary
containing a white dwarf accreting material from the wind of the red giant. 

The X-ray spectrum of 4~Dra is sometimes dominated by
partially-ionised photoelectric absorption, presumably due to the wind
of the red giant. We note that X-ray monitoring of such systems would
provide a powerful probe of the wind and mass-loss rate of the giant,
and would allow a detailed test of wind accretion models. 
\end{abstract}


\section{Introduction}
%%%%%%%%%%%%%%%%%%%%%%
\scite{Reimers85} 
reports
the discovery
of an ultraviolet companion to the M3\,III giant 4~Draconis. 
International Ultraviolet Explorer (IUE) observations show its 
spectrum is similar to that of high-accretion-rate cataclysmic variables, 
with a slowly decreasing 
continuum in the range 3000--1500\,\AA ~which then rises steeply to 
shorter wavelengths. There are strong and broad high-excitation emission 
lines (also typical of cataclysmic variables) and some narrow 
low-excitation emission lines which Reimers attributes to the ionised wind of 
the giant.

More detailed ultraviolet and optical-radial-velocity measurements are 
presented by \scite{Reimers88}.
They determined the orbit of the giant 
and claimed that the ultraviolet flux is modulated at a period of 4\,h. 
Based largely on this period, they conclude 
that the ultraviolet companion is most likely an AM~Her-type 
cataclysmic variable. 

\scite{Eggleton89} point out that the orbit of the wide pair in 4~Draconis 
($\rm P_{orb}$=1703\,d) severely limits the size of the progenitor
of a cataclysmic variable and places unique constraints on its evolution.
Without constraints on even the inclinations of the two binary orbits they 
argue that any cataclysmic variable must have evolved from a progenitor
with initial $\rm P_{\sc orb}\leq$100\,d. 
However Eggleton et al.\   also point out that the 
identification as an AM~Her system is uncertain,
and 
that an isolated white dwarf may explain the observations equally well. 
In this picture the white dwarf would be accreting from the wind of the giant 
and the four hour period would be its spin period. This requires a magnetic
field on the white dwarf sufficient to funnel the accretion flow onto its 
poles.

In this paper we present the first X-ray observations of 4~Draconis, and 
discuss the nature of the system. 

\section{Observations}
\label{sec-obs}
4~Draconis was observed four times with ROSAT \cite{Trumper83}:
once during the ROSAT all-sky survey (RASS) with the 
position sensitive proportional counter (PSPC, \ncite{Pfeffermann87});
twice during the pointed phase of the mission with the PSPC, 
once as the target and once serendipitously; 
and once with the high-resolution imager (HRI, \ncite{Zombeck95}). 
A log of the pointed 
observations is presented in Table\,\ref{tab-log}. 
4~Draconis was detected in all four observations (see Sect.\,\ref{sec-det})
and lightcurves and spectra were extracted from circular regions of
radius 1.3, 6.0, and 0.6\,arcmin for the 1991, 1993 and 1996
pointed observations respectively. Background rates were estimated using 
large nearby regions free from obvious point sources. 
Raw count rates are presented in Table\,\ref{tab-centroid}, as well as
our best estimates of the equivalent on-axis PSPC count rate. The
serendipitous off-axis observation has been corrected for vignetting
and for source counts lost outside the selection radius (total factor
1.91). The RASS count rate \cite{Huensch98} 
has been corrected using an intermediate factor (1.45). 
The HRI count rate has been corrected using a factor 4.1 derived from PIMMS 
\cite{Mukai-PIMMS} and assuming our best fitting spectrum to the 1993
ROSAT spectrum of 4~Dra (see Sect.\,\ref{sec-spec}).

\begin{table}
\begin{center}
\caption{Log of ROSAT pointed observations of 4~Draconis.}
\label{tab-log}
\begin{tabular}{ccccr}
Obs.\  date & Instr. & Sequence no.\  & Off-axis$^a$ & Exp.$^b$ \\\hline
4--5 Apr 1991 & PSPC & 300034 & 0 & 15.0  \\
5--6 Jun 1993 & PSPC & 701225 & 51 &  5.7 \\
8--20 Oct 1996 & HRI  & 300492 & 0 & 41.8 \\
\end{tabular}
\end{center}
{\small $^a$ Off-axis angle [arcmin]. }\\
{\small $^b$ Exposure [ks]. }\\
\end{table}

\section{Results}
\subsection{X-ray detection}
\label{sec-det}
An X-ray source is detected at the position of 4~Draconis in all three ROSAT 
observations. Table\,\ref{tab-centroid} shows the results of fitting for the 
centroid of the spatial count distribution. In each case the difference in 
position between the count centroid and 4~Draconis is less than or equal to 
the half width at half maximum (HWHM) of the distribution. 
The probability of chance alignment is small. For example, the probability of 
a single source lying so precisely at the centre of the HRI detector is 
$\sim6\times10^{-5}$. There are only about ten sources detected in the image, 
so we are confident that the probability of chance alignment is $<10^{-3}$ and
that the detected source is 4~Draconis.
The count rates presented in Table\,\ref{tab-centroid} show that it is
highly variable.

\begin{table}
\begin{center}
\caption{Count rates and results of centroid fitting for the
ROSAT observations of 4~Draconis. 
$\Delta \theta$ is the angular separation between the position of 4~Draconis 
and the fitted centroid of the spatial count distribution. HWHM is the 
half width at half maximum of the count distribution, also expressed as an 
angle.}
\label{tab-centroid}
\begin{tabular}{clcccc}
 & & & & \multicolumn{2}{c}{Count rate [s$^{-1}$]} \\
Obs.\  date & Inst. &$\Delta \theta^a$& HWHM$^a$ & raw & corrected$^b$ \\\hline
Nov 1990 &RASS& - & - & 0.033$\pm$0.008 & 0.048 \\
Apr 1991 &PSPC& 0.26& 0.35 & 0.011$\pm$0.001 & 0.011 \\
Jun 1993 &PSPC& 0.6\phantom{0} & 3.1\phantom{0}& 0.371$\pm$0.008 & 0.707 \\
Oct 1996 &HRI& 0.15 & 0.14 & 0.059$\pm$0.001 & 0.172 \\
\end{tabular}
\end{center}
{\small $^a$ [arcmin].} \\
{\small $^b$ Corrected to PSPC on-axis response.} \\
\end{table}

\subsection{X-ray spectroscopy}
\label{sec-spec}

\begin{figure}
\begin{center}
\includegraphics[width=8.4cm]{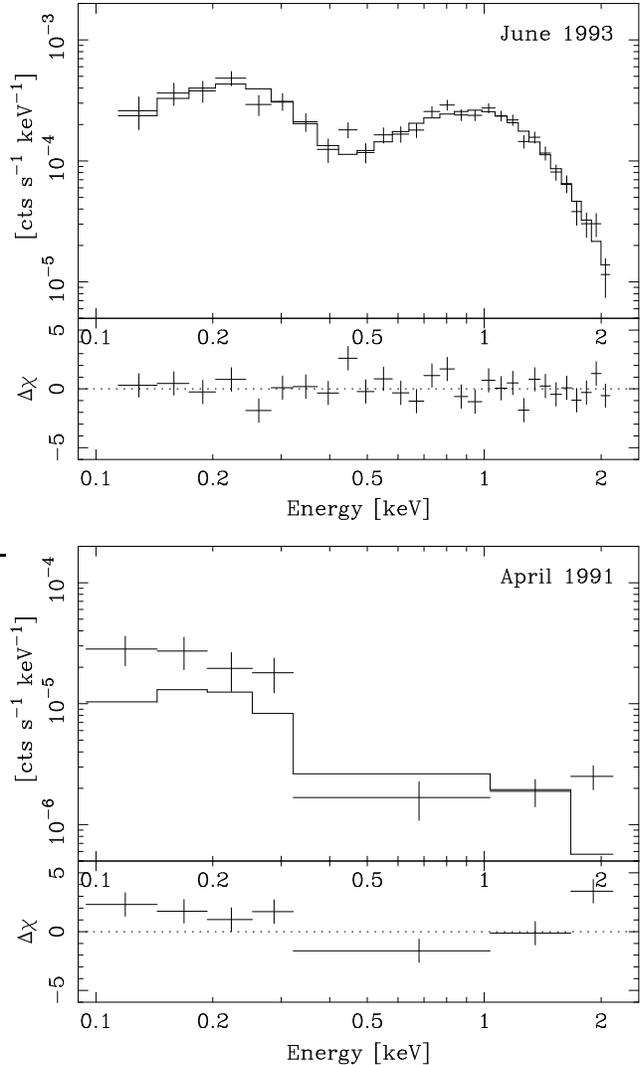}
\caption{\label{fig-spec} 
ROSAT PSPC spectra of 4~Draconis. Both spectra have been fit with a 
optically-thin thermal plasma model. The lower panel in each case
shows the fit residuals scaled by the error on each data point. }
\end{center}
\end{figure}

\begin{figure}
\begin{center}
\includegraphics{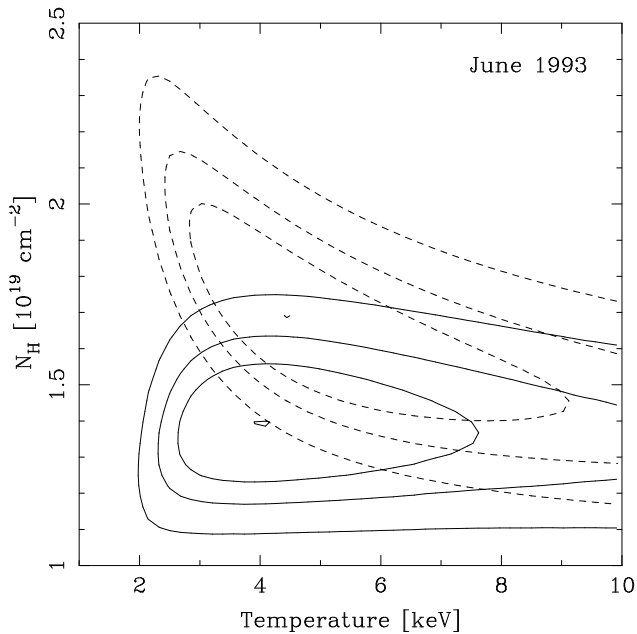}
\caption{\label{fig-grid} 
Contour plots of allowed parameter values for our fits to the 1993
PSPC spectrum of 4~Dra. Solid contours show the constraints on a
bremsstrahlung model. Dotted contours show the constraints on the {\it
mekal} plasma model. Contours represent 68, 90 and 99 per cent
confidence for two interesting parameters 
($\Delta \chi^2$=2.3,4.61,9.21 respectively).
}
\end{center}
\end{figure}

\begin{figure}
\begin{center}
\includegraphics{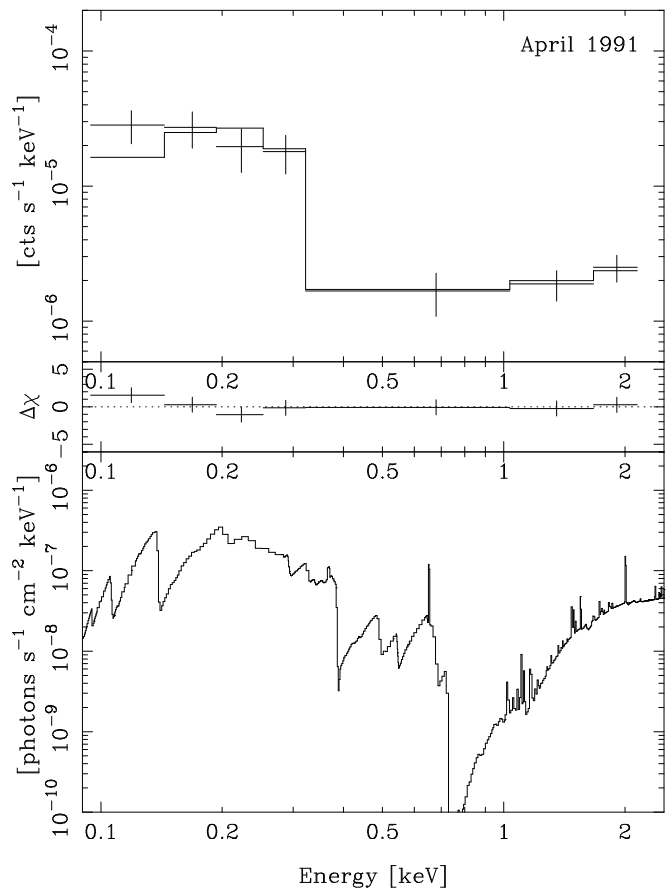}
\caption{\label{fig-absori} 
The 1991 ROSAT PSPC spectrum of 4~Draconis fitted with an ionised absorber 
model. The top panel shows the observed spectrum, the middle panel shows the 
fit residuals and the lower panel shows the model spectrum.
}
\end{center}
\end{figure}

Figure\,\ref{fig-spec} shows the spectra of 4~Draconis extracted from the two 
PSPC pointed observations. 
They have been fit with an optically-thin thermal plasma 
model \cite{Mewe85,Kaastra93}. 
The off-axis 1993 observation is well fit with this model, 
with a temperature of 4.4$\pm^{2.6}_{0.8}$\,keV 
and absorbing column density of $\rm (1.5\pm0.2)\times10^{19}\,cm^{-2}$ 
(reduced $\chi^2$=1.05 with 26 degrees of freedom). 
Figure\,\ref{fig-grid} shows the constraints on these parameters. 
In contrast, the 1991 observation 
is very poorly fit with this model (reduced $\chi^2$=6.7 with 4 d.o.f.). 

The unabsorbed 0.1--2\,keV flux for the fit to the 1993 spectrum is 
$9\times10^{-12}\rm\,erg\,s^{-1}\,cm^{-2}$. Including a bolometric correction 
factor of 2.4, this corresponds to a luminosity of 
$6\times10^{31}\rm\,erg\,s^{-1}$ at the HIPPARCOS distance of 4~Draconis 
(178\,pc; \ncite{Perryman97}).

The fit residuals for the 1991 spectrum 
show a minimum around 1\,keV (Fig.\,\ref{fig-spec}) 
which corresponds to the maximum of the
effective area of the PSPC and so must represent a true minimum in the X-ray 
flux. No physically-plausible pure-emission model can reproduce this minimum, 
but a partially-ionised absorber does so naturally. 
Photoelectric absorption by 
cold cosmic-abundance material increases to low energies, but soft photons can
leak through 
if low energy edges have been removed through ionisation.
We find that such a model readily reproduces the observed 1991 spectrum, and 
we plot a typical fit in Fig.\,\ref{fig-absori} 
(with hydrogen column density of $4\times10^{23}\rm\,cm^{-2}$ and
ionisation parameter $\xi$=5.8). 
Unfortunately the combination of low spectral 
resolution and low signal-to-noise prevents a unique fit to an ionised 
absorption model and we cannot well constrain the properties of the absorbing 
medium. However, the wind of the red giant star is an obvious candidate 
absorber and the X-ray source itself may supply the photo-ionising
flux. 
The difference between the 1991 and 1993 observations may be explained in part
or entirely by a changing column density and/or ionisation fraction
along our line of sight through the wind. 

\subsection{X-ray time-series analysis}
\begin{figure*}
\begin{center}
\includegraphics[width=17.7cm]{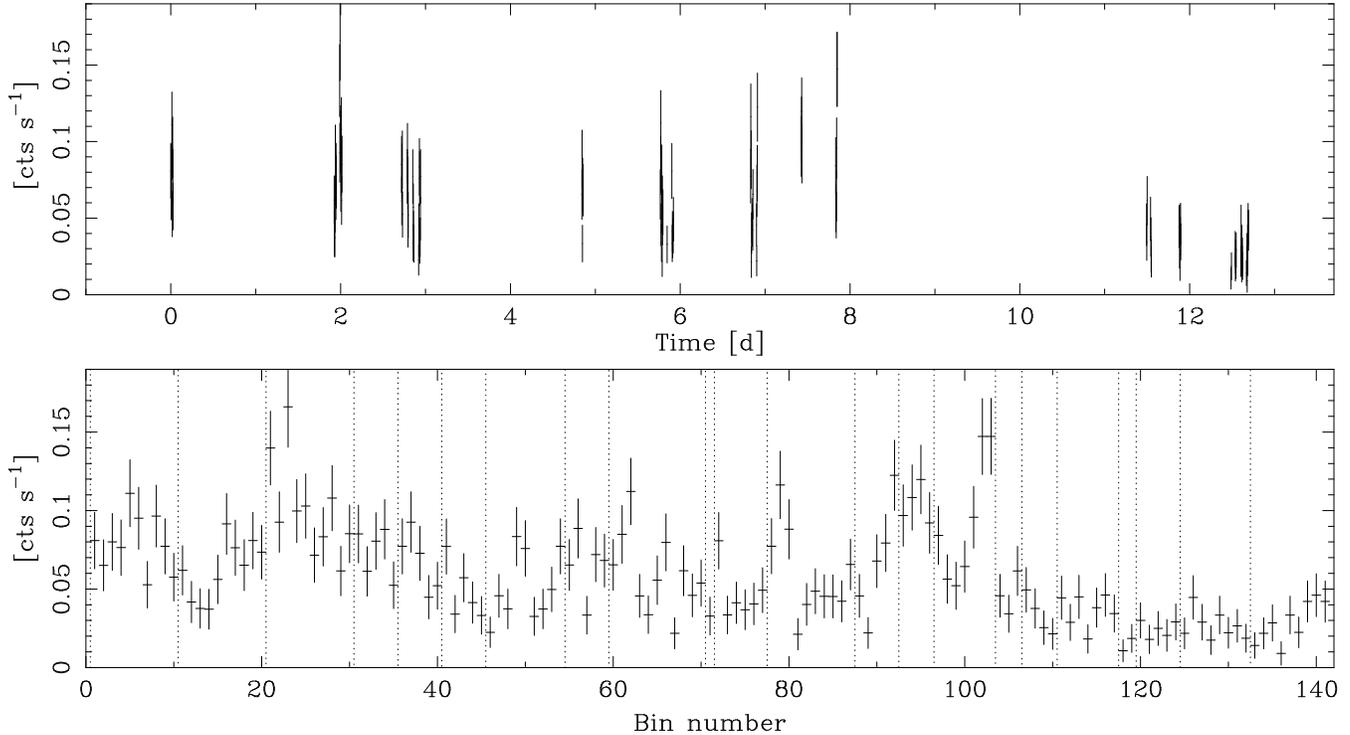}
\caption{\label{fig-lc}
The 1996 ROSAT HRI lightcurve of 4~Draconis binned into 256\,s bins. 
For clarity, data gaps have been omitted in the lower panel (dotted lines 
indicating breaks in the time axis). The top panel shows the full lightcurve. 
}
\end{center}
\end{figure*}

\begin{figure}
\begin{center}
\includegraphics[width=8.5cm]{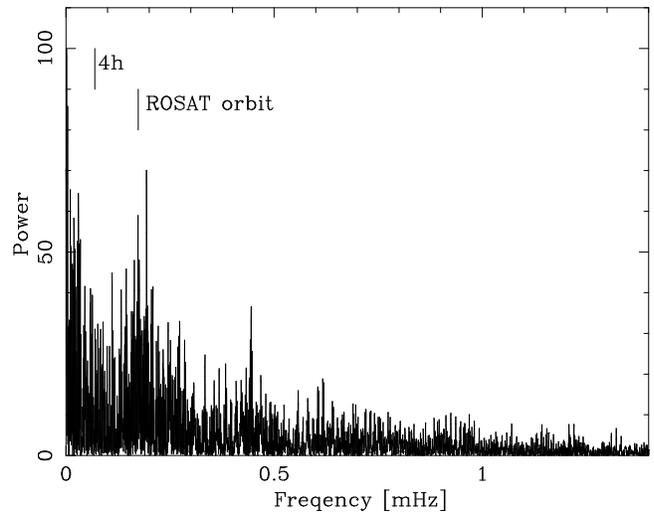}
\caption{\label{fig-pspe} 
The power spectrum of the 1996 ROSAT HRI lightcurve of 4~Draconis. 
}
\end{center}
\end{figure}

The lightcurves from all three ROSAT observations reveal variability on short 
timescales. The lightcurve from the 42\,ks HRI observation is presented in
Figure\,\ref{fig-lc}.
It shows strong variability on timescales between minutes and days, but 
there is no evidence for the four-hour periodic modulation claimed
by \scite{Reimers88} from IUE observations.
Figure\,\ref{fig-pspe} shows the power spectrum of the HRI lightcurve,
in which no obvious periodic signal is apparent. There is excess power close 
to the ROSAT orbital period (96\,min), 
with the strongest peak at a slightly shorter period (86\,min), and there 
is another suggestive peak at 37.5\,min, but neither are sufficiently strong to
represent a conclusive detection of periodic modulation. 

\section{Discussion}
\subsection{The nature of 4~Draconis}
Previous ROSAT observations have shown that red giants are not substantial 
X-ray emittors. 
Only one
late-type giant was detected in the ROSAT all-sky survey
\cite{Haisch91,Haisch92,Huensch96}, 
and pointed observations placed an extremely tight upper limit 
of $3\times10^{25}\rm\,erg\,s^{-1}$
on the X-ray flux of the K1\,III red giant Arcturus \cite{Ayers91}. 
Thus we can be confident that the X-ray emission from 4~Draconis reported in
this paper originates on the ultraviolet companion, 4~Dra\,B.

Our ROSAT observations are consistent with this secondary containing an 
accreting white dwarf. 
The $\sim$5\,keV temperature of the optically-thin X-ray spectrum is 
characteristic of non-magnetic cataclysmic variables
\egcite{Wheatley96} and of the ``bombardment solution'' for radial 
accretion onto a white dwarf \egcite{Woelk95}. The bombardment
solution 
applies when the mass accretion rate per unit area is too low for a 
stand-off shock to form 
($\rm\dot{m}<10^{-1}\,g\,s^{-1}\,cm^{-2}$).
Our measured luminosity of 
$6\times10^{31}\rm\,erg\,s^{-1}$ 
implies an accretion rate of 
$0.24-1.8\times10^{15}\rm\,g\,s^{-1}$
for white dwarf masses in the range $0.3-1.0\rm\,M_{\sun}$. 
For the bombardment solution to apply this accretion rate must be
spread over an area of at least 
$0.24-1.8\times10^{16}\rm\,cm^2$, 
although this is a small fraction of the surface area of even a
massive white dwarf. 

Although our observations are consistent with the presence of an
accreting white dwarf, they do not support the presence of an AM~Her
system. First, the ROSAT spectra of AM~Hers are
typically dominated by intense optically-thick soft emission, 
with characteristic temperatures of $\sim$20\,eV. We can rule out the
presence of such a component in the 1993 spectrum of 4~Dra\,B 
(Fig.\,\ref{fig-spec}). 
Second, 
it is
clear from the HRI lightcurve (Fig.\,\ref{fig-lc}) that the X-ray 
emission is not strongly modulated at a period of 1--8\,h, 
as it is for every known high-state AM~Her system and most other
magnetic cataclysmic variables. 

AM~Her systems have shown spectra much like that of 4~Draconis during
low accretion rate states \egcite{Ramsay95}, but our measured
luminosity is rather high for a low-state AM~Her, and we believe the lack
of an X-ray orbital periodicity alone is sufficient evidence to rule out the
presence of an AM~Her in the 4~Draconis system.

Non-magnetic cataclysmic variables (e.g.\   dwarf novae) usually have 
no optically-thick component in the ROSAT bandpass \egcite{Wheatley96},
have characteristic X-ray temperatures lower than AM~Hers, and do not
exhibit strong orbital X-ray modulation. 
Therefore we cannot rule out the presence of a 
non-magnetic cataclysmic variable.
However, the original case for the presence of a cataclysmic variable was based
upon the claimed detection of a 4\,h ultraviolet period 
\cite{Reimers88}. Reviewing the lightcurve in Fig.\,3 of Reimers et al.\  
we believe the case for a periodic modulation is not strong. Also,
more recent HST observations do not support the presence of a 4\,h
period (B.\  Gaensicke, private communication). Thus, in
the face of evidence clearly supporting the presence of an accreting white 
dwarf, but none requiring the presence of a third star, we conclude that the 
companion to 4~Draconis is most-likely a single white dwarf. 
The 4~Draconis system is then most-likely a symbiotic binary. 

\subsection{X-ray spectra of symbiotic stars}
We note that the absorbed 1991 X-ray spectrum of 4~Dra is strikingly
similar to that of the symbiotic star CH~Cyg measured with ASCA
\cite{Ezuka98}. \scite{Muerset97} and \scite{Ezuka98} interpret the
X-ray spectrum of CH~Cyg as a combination of accretion emission
(hard X-rays) and the colliding winds of the two stars (soft
X-rays). However, \scite{Wheatley01} and Wheatley (in preparation) 
reanalysed the ASCA spectrum and found that it could be interpreted
instead as a single accretion-driven emission component viewed through 
a partially-ionised absorber. Our discovery of the same behaviour in 4~Dra
shows that there is a distinct subclass of symbiotic stars in which
the X-ray spectrum is dominated by complex absorption of a hard X-ray
accretion-driven continuum. 4~Dra and CH~Cyg seem
to be distinct from the other symbiotic stars studied by
\scite{Muerset97} which have much softer X-ray emission. 

\subsection{Absorption and orbital phase}
Our results show a marked decrease in absorption between the 1991 and
1993 ROSAT observations of 4~Dra. If the absorption is due to the wind of the
red giant, then the change in absorption may be related to our
changing line of sight through the wind to the X-ray source. 

\scite{Reimers88} studied the radial velocity variations of the red
giant and measured its orbital elements. They find a mildly eccentric 
orbit, $e=0.3$, with a 5\,yr period. The orbital phases of the two
ROSAT PSPC observations, relative to periastron, are $0.22\pm0.03$ in
1991 and $0.68\pm0.03$ in 1993. It can be seen that the high
absorption spectrum in 1991 was taken slightly closer to periastron,
and the unabsorbed 1993 spectrum was taken slightly closer to
apastron. Probably more importantly, the orbit is orientated such that
the phase of the 1993 spectrum corresponds to a time when the X-ray
source is in front of the red giant. This may explain the reduced
absorption in this spectrum. 

\begin{figure}
\begin{center}
\includegraphics[width=8.5cm]{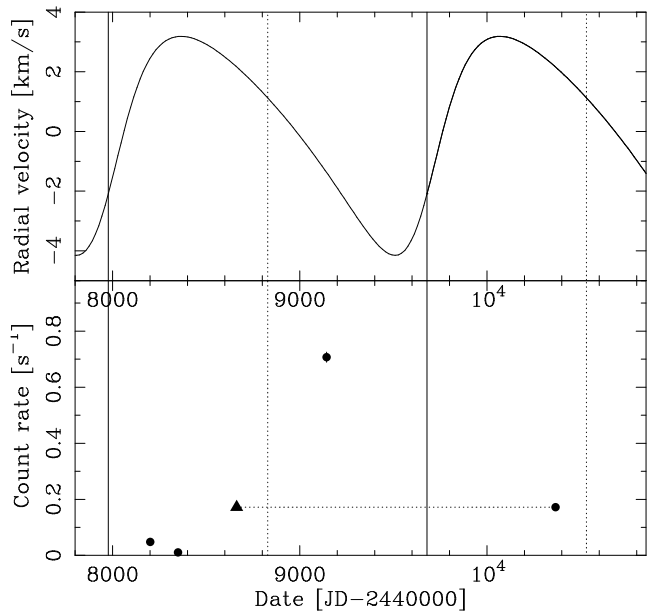}
\caption{\label{fig-long} 
Long-term X-ray lightcurve of 4~Draconis (bottom panel) using estimated
equivalent PSPC on-axis count rates for all four ROSAT observations 
(see Table\,\ref{tab-centroid} and Sect.\,\ref{sec-obs}).
The 1996 HRI observation is plotted twice in order to show its
relative orbital phase with respect to the other observations (triangle).
The top panel shows the radial velocity 
of the giant star as determined by Reimers, Griffin \& Brown (1988).
The solid vertical lines represent periastron, and the dotted
lines represent apastron. 
The orbital phase of the bright 1993
observation corresponds to a time when the X-ray source is in front of
the giant. 
}
\end{center}
\end{figure}

Figure\,\ref{fig-long} shows the long term X-ray lightcurve of 4~Dra
using the estimated PSPC on-axis count rates for all four observations
(see Table\,\ref{tab-centroid} and Sect.\,\ref{sec-obs}). It can be
seen that the four ROSAT observations are consistent with a smooth
variation in count rate with orbital phase. 

Further observations are required in order to test whether the X-ray 
absorption in the spectrum of 4~Dra\,B really varies smoothly with orbital
phase. We note, however, that a set of X-ray observations covering all
orbital phases would provide a powerful probe of the wind and
mass-loss rate of the red giant star, and would allow a detailed test of
wind accretion models. 

\subsection{Wind accretion}
In order to test our conclusion that 4~Dra\,B is most likely a single
white dwarf accreting from the wind of the giant, we estimate the
expected wind accretion rate in such a system. For the purpose of this
order-of-magnitude estimate we assume the white dwarf is
non-magnetic. A rotating magnetic field might act to reduce this
accretion rate through the propeller effect. 

The case of a white
dwarf accreting from the wind of a red giant was treated by
\scite{Livio84}. Using their Equation\,3 we find a wind accretion
luminosity of $L_{acc}\sim10^{34}\,\rm erg\,s^{-1}$ using the values:
stellar separation $a\sim4\,{\rm AU}=6\times10^{13}\rm\,cm$ \cite{Reimers88}, 
wind velocity $v_w\sim10\,\rm km\,s^{-1}$ \cite{Reimers89},
orbital velocity of the white dwarf
$v_{0}\sim30\,\rm km\,s^{-1}$, 
and a wind mass loss rate 
$\dot{M}_w\sim10^{-8}\,M_{\sun}\,\rm yr^{-1}$.

This luminosity is larger than both our measured X-ray luminosity, 
$6\times10^{31}\rm\,erg\,s^{-1}$, and the ultraviolet luminosity
measured by \scite{Reimers85}, $1\times10^{33}\rm\,erg\,s^{-1}$
(corrected to the HIPPARCOS distance, 178\,pc). Many of the values used to
estimate the wind accretion rate are uncertain, but it is clear that
direct wind accretion {\em is} a viable energy source for the
high-energy emission of 4~Dra\,B.

We note that the ratio of X-ray to ultraviolet luminosities is similar
to that measured for the disc-accreting white dwarfs in dwarf novae 
\egcite{Wheatley96}.

The presence or otherwise of an accretion disc in a wind accretor
depends sensitively on the relative velocity of the wind and accreting
object \cite{Livio84}. Given the uncertainty in this quantity we
cannot be sure whether or not an accretion disc is present around
4~Dra\,B. However, \scite{Reimers85} and \scite{Reimers88} argue that
the broad high-excitation emission lines of 4~Dra are more like those
of AM~Her systems than disc accreting dwarf dwarfs. 
This may point to the absence of an accretion disc in 4~Dra, with
these broad high-excitation lines arising in the X-ray-illuminated,  
radial, accretion flow: a geometry much like that found in AM Her 
systems. 

\section{Conclusions}
We present the discovery of X-ray emission from the symbiotic system
4~Draconis. The X-ray flux is highly variable on timescales from
minutes to years. X-ray spectroscopy shows the spectrum is sometimes
dominated by strong absorption by partially ionised material, probably
the wind of the red giant. When free from absorption the spectrum is 
consistent with bremsstrahlung emission at a temperature around
5\,keV. We conclude that these data are consistent with the presence of an 
accreting white dwarf, but that the lack of periodic X-ray modulation
rules out the previously proposed identification of the hot companion 
as an AM~Her system \cite{Reimers88}. 
Instead we conclude that the companion is most 
likely a single white dwarf accreting from the wind of the giant. 
Consequently the evolutionary constraints derived by \scite{Eggleton89}
need not apply. Finally, we show that wind accretion is a viable energy 
source in this system.

\section{Acknowledgments}
We thank Boris Gaensicke for providing information on his HST
observations in advance of publication. 
Astrophysics research at the University of Leicester is supported by
PPARC rolling grants. 
ROSAT data were extracted from the 
Leicester Database and Archive Service (LEDAS) at the University of Leicester.

\bibliographystyle{mnras3}
\bibliography{mn_abbrev2,refs}

\end{document}